\documentclass[pra,aps,superscriptaddress,twocolumn]{revtex4}
\usepackage{amsmath,amssymb}
\usepackage{graphicx} 
\usepackage{color}
\usepackage{braket} 
\usepackage{amsthm}
\usepackage{tikz}
\usepackage{caption,pdfpages}
\usepackage{pgfplotstable} 
\usepackage{pgfplots} 
\usetikzlibrary{decorations.markings}  
\tikzset{
  font={\fontsize{7 pt}{12}\selectfont}}
\newtheorem{theorem}{Theorem}

\newtheorem{lemma}[theorem]{Lemma}
 
\newtheorem{definition}{Definition}

\begin{document} 

\author{Karthik S. Joshi}
\affiliation{Poornaprajna Institute of Scientific Research, Bengaluru, India.}
\author{S. K. Srivatsa}
\affiliation{Dept. of Electrical and Electronics Engineering, PES University, Bengaluru, India.}
\affiliation{Poornaprajna Institute of Scientific Research, Bengaluru, India.}
\author{R. Srikanth}
\email{srik@poornnaprajna.org}
\affiliation{Poornaprajna Institute of Scientific Research, Bengaluru, India.}

\title{Path integral approach to one-dimensional discrete-time quantum walk}

\begin{abstract}
Discrete-time quantum  walk in  one-dimension is studied  from a  path-integral perspective. This enables derivation of a closed-form expression  for amplitudes corresponding to any coin-position basis of  the  state vector  of the  quantum  walker at  an arbitrary step of the walk. This provides a new approach to the foundations and applications of quantum walks.
\end{abstract}

\maketitle

\section{Introduction}
An approach to quantum mechanics via the generalization of the action principle in classical mechanics was established by Feynman \cite{feynman1948space}, known as the \textit{path integral} formulation. It assumes that a particle can take all possible paths to travel between two fixed points, say $O$, the origin, and $T$, the terminal point respectively. A probability amplitude is assigned to each path the particle can take and the propagator is obtained by taking the sum of all amplitudes corresponding to all paths for the particle  to propagate  from  $O$ to $T$.

Let $\mathcal{C}$ be a path from $O$ to $T$, with an assigned probability amplitude $P(\mathcal{C})$, which in general is a complex number. The probability that the particle will take the path $\mathcal{C}$ is $|P(\mathcal{C})|^2$. The joint probability for the particle to take paths $\mathcal{C}_1$ and $\mathcal{C}_2$ is then $|P(\mathcal{C}_1)+P(\mathcal{C}_2)|^2$. Notice that the rule for obtaining joint probability deviates from its classical counterpart by an interference term of the form $P(\mathcal{C}_1)P^*(\mathcal{C}_2)+P^*(\mathcal{C}_1)P(\mathcal{C}_2)$. Further, in a path $\mathcal{C}$ made of two sub - paths say $\mathcal{C}_1$ and $\mathcal{C}_2$ meeting at an intermediate point $M$, the probability amplitude $P(\mathcal{C})=P(\mathcal{C}_1)P(\mathcal{C}_2)$.      

The path integral approach has been applied to study various physical processes, including the quantum dynamics involving scalar potentials with singularities which were not amenable to the standard operator approach of Schrödinger \cite{nelson1964feynman}. It has also been extensively applied in the study of quantum field theory particularly for quantization, gauge fixing and phenomenon concerning elementary particles \cite{ashok2006field}. The problem of decoherence, central for realization of a quantum computer has been discussed via the path integral approach in \cite{albeverio2007rigorous}. 

A candidate system to study via Feynman approach is the quantum walk (QW). The QW is a process wherein the evolution of the walker is driven by a fixed unitary operator.
QW process begins with the walker at the origin, say, moving backward or forward conditioned on the outcome of a (quantum) coin toss. This process is repeated $n$ - times for a $n$ - step QW. 
Broad classification of QW include discrete time QW (DTQW) and continuous time QW. Many variants of QW have been studied; alternating \cite{franco2011mimicking}, coinless \cite{patel2005quantum} and staggered model \cite{portugal2016staggered}. QW has also been studied in the presence of noise, for instance \cite{srikanth2010quantumness}.  A detailed review of quantum walk is presented in \cite{rev}.  

Realization of QW have been carried out in a variety of physical systems and have been discussed in detail in \cite{PIQW}. Schemes to implement QW have been proposed; ion \cite{Trav} and optical traps \cite{Chandru06}, quantum circuits \cite{Kosik} and cavity QED \cite{Agarwal} to name a few.

It has been shown that QW serves as an effective tool to design quantum algorithms.  Algorithmic applications of quantum walk include \textit{element search} \cite{ne} and \textit{triangle finding} (finding a triangle on an undirected graph and also indicate if the graph is triangle free) \cite{Fre} among others. Also, it has been shown that any quantum computation process
can be implemented by performing quantum walk in both continuous and discrete cases \cite{childs2009universal} \cite{lovett2010universal}. This powerful result only adds more thrust to the study of quantum walk. 

Further, close ties of the QW framework and quantum field theory has recently come to fore. Among others, QW as a model to perform quantum simulation of relativistic particles has been studied \cite{pablo2014the}. With the demonstration of implementation of QW using a Dirac like Hamiltonian \cite{chandrashekar2013two}, modelling the phenomena of neutrino oscillations \cite{mallick2017neutrino}, etc the study of QW has only gained further impetus.

While the rich and varied applications of QW indicates the success of the framework, analytical results are but few. This can be attributed to the many interferences occurring simultaneously, which seemingly makes it analytically intractable. However, the maxima of the probability distribution arising from the QW  process was obtained analytically in \cite{nayak2000quantum}.    In this work, we study QW in one dimension via the Feynman path integral approach, obtaining in the process a closed form expression for the probability amplitude for an arbitrary time step in the case of a one dimensional DTQW.

\section{Feynman path approach to QW}
We briefly present the Discrete Quantum walk scheme for completeness.  
Consider the walker to be initially in the state
\begin{align}
\label{eq: inistate}
\ket{\psi(t=0)}=(\alpha\ket{0} +e^{i\phi}\beta\ket{1})\otimes \ket{k}.
\end{align} 

We assume $\alpha$ and $\beta$ to be real numbers without loss of generality. The Hilbert space of the walker is $\mathcal{H}=\mathcal{H}_{coin} \otimes \mathcal{H}_{position}$. One-step evolution of the walker is achieved by applying a unitary operator $U \equiv C \otimes S$ on $|\psi(0)\rangle$, with 
coin operator $C$, given by:
\begin{equation}
C =
\begin{pmatrix}
\cos\theta & \sin\theta \\
\sin\theta & -\cos\theta
\end{pmatrix},
\end{equation}
and the shift operator $S$ given by:
\begin{equation}
S=\Sigma_k \ket{0}\bra{0} \otimes \ket{k-1} \bra{k} +\Sigma_k \ket{1}\bra{1}  \otimes \ket{k+1}\bra{k}.
\end{equation}
Application of $U^n$ carries out a $n \ $- step QW. 
Evolving the initial state Eq.(\ref{eq: inistate}) by one step we obtain 
\begin{align}
\ket{\psi(1)}&=\alpha\cos\theta\ket{0,k-1}+\alpha\sin\theta \ket{1,k+1} \nonumber \\
 &+e^{i\phi}\beta\sin\theta\ket{0,k-1}-e^{i\phi} \beta \cos\theta \ket{1,k+1}.
\label{eq: onestp}
\end{align}

\begin{center}
\begin{figure}[h] 
\scalebox{0.8} {\includegraphics[trim={1.8cm 23cm 2cm 1.8cm},clip]{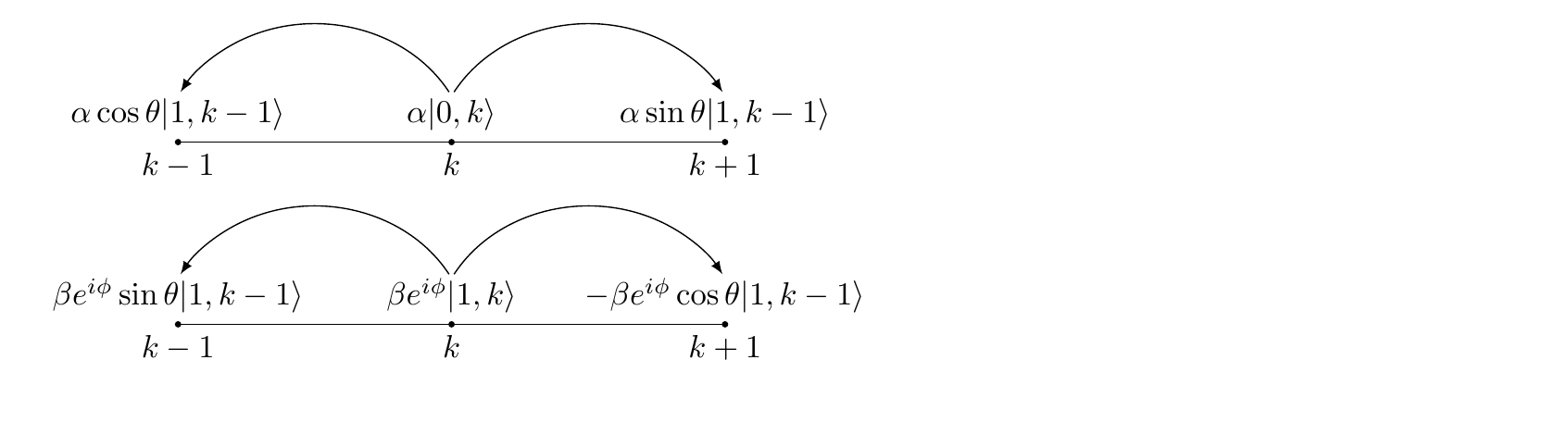}} 
\caption{Figure shows all possible one step transitions and the corresponding probability amplitudes of the walker starting with initial states $\alpha\ket{0,k}$ and $\beta e^{i\phi}\ket{1,k}$}
\label{fig:path}
\end{figure}
\end{center}

The basis vectors of $\ket{\psi(1)}$ represent the final state of the walker. Let the transitions $\ket{k} \rightarrow \ket{k+1}$ and $\ket{k} \rightarrow \ket{k-1}$ be labelled as \textit{forward} ($F$) and \textit{backward} ($B$) transition respectively. All possible one-step transitions are deduced from Eq.(\ref{eq: onestp}) and are illustrated in Fig. (\ref{fig:path}).

Note that the probability amplitudes are independent of the instantaneous position of the walker, $\ket{k}$ and the step number, $t$.     

Two steps of the QW can be obtained by combining one-step transitions. For instance, a possible two step evolution can be obtained by combining $B$ and $F$ one-step transitions- a $BF$ transition, which indicates that the walker underwent a $\ket{k} \rightarrow \ket{k-1}$ transition at time $t=1$ followed by a $\ket{k-1} \rightarrow \ket{k}$ transition at time $t=2$. Similarly, an $n \ $- step walk can be represented by a string of $F$s and $B$s of length $n$, which prompts the following definition:

\begin{definition}[Feynman string] 
Any $n$-bit  string $S  \in \{B, F\}^{\times n}$ represents the $n-1$ (one step) transitions in a Feynman path of an $n \ $ - step QW.
\end{definition}

Strings starting with a $B$ are called $B$ - strings, likewise, strings starting with $F$ are called  $F$ - strings. Given a string $S$, its \textit{dual} $\bar{S}$, is obtained by interchanging $F$s and $B$s. Evidently, the dual of a $B$-string is an $F$-string and vice-versa.

For any path starting at the origin and terminating at position $\ket{x}$ (say), the corresponding Feynman string $S$ has $\frac{n+|x|}{2}$ $F$s and $\frac{n-|x|}{2}$ $B$s. Hence, the number of strings $N$ representing paths which start at the origin and terminate at $\ket{x}$ is given by the number of distinct permutations of $S$. Therefore:
\begin{align}
N= \frac{n!}{\left(\frac{n \pm |x|}{2}\right) ! \ \left(\frac{n \mp |x|}{2}\right) !} = {n \choose \frac{n \pm |x|}{2}}. 
\label{eq:TotFP}
\end{align}   

The walker has a certain initial coin state as shown in Eq.(\ref{eq: inistate}). This obviously affects the evolution of the walker, as coin states $\ket{0}$ and $\ket{1}$ effect $B$ and $F$ translations respectively. This gives rise to the mapping $B \rightarrow \ket{0}$ and $F \rightarrow \ket{1}$. Each of the above $N$ strings are thus concatenated with a $B$ or an $F$, effectively giving rise to strings $B \ast S$ and $F \ast S$ for all $S$, leading to the following definition:

\begin{definition}[Feynman extended string]
An $(n+1)$-bit string, denoted $S^+$, consisting  of a Feynman
string preceded by the initial coin state of the walker $B$ $(\ket{0})$ or $F$ $(\ket{1})$.
\end{definition}

Note that each of the extended strings $S^+$ of an $n$ - step walk has $n$ one step transitions. Furthermore, $B \ast S$ and $F \ast S$ for all $N$ number of strings $S$, will result in $2N$ strings.
   
In the Feynman framework, each of the $2N$ paths are associated with a probability amplitude. Let the $m^{th}$ path string have a  probability amplitude $P(S_{m})$. The net probability amplitude, $P(x)$, for the walker to be found at position $\ket{x}$ is the sum of all these amplitudes:
\begin{align}
P(x) = \sum_{m=1}^{2N}P(S_{m}).
\end{align}

On the other hand, for a given path represented by an $n+1$ - bit string $S$, the corresponding probability amplitude $P(S)$ is given by the product of all the $n$ one step probability amplitudes $P(k)$:

\begin{align}
P(S)=\prod_{k=1} ^{n} P(k). 
\end{align} 

The discussion presented in the remainder of the article pertains to both Feynman string and its extended counterpart. They will be collectively referred to as Feynman strings unless otherwise specified.   

Note that each one step transition introduces a $\sin\theta$ or a $\pm \cos\theta$ factor to the probability amplitude, therefore it is proportional to $\cos^i\theta \sin^j\theta$, with $i+j=n$. 

\begin{definition}[Switch] 
In  any  $n$-bit  Feynman string,  $S \in  \{B,  F\}^{\times  n}$,  
\textit{switch} is  defined as a  $F \leftrightarrow B$ transition in the string.
\end{definition}  

It is easily seen from Fig. (\ref{fig:path}) that number of switches fixes the degree of $\sin\theta$ in the probability amplitude. Evidently, there are multiple strings, say $\eta (n, x, j)$, with same number of switches, which needs to be accounted.
    
The full path string of length $n$ and terminating at position $|x|$ 
has $N_{F}(n,x) = \frac{n+|x|}{2}$ dominant (say, forward) shifts and
$N_{B}(n,x) = \frac{n-|x|}{2}$ minority (say, backward) 
shifts. Note that for a fixed time step $t$, $n$ and $|x|$ are both even or both odd.

The number of strings, for a fixed $j$ switches, is equivalent to the problem of filling 
(i) $N_{F}(n,x)$ balls into $\mu\equiv\lfloor (j+1)/2\rfloor $ urns \textit{and} of
$N_{B}(n,x)$ balls into $\nu \equiv\lceil (j+1)/2\rceil$ urns; plus
(ii) $N_{F}(n,x)$ balls into $\nu$ urns \textit{and} of $N_{B}(n,x)$ balls into $\mu$ urns; 

At the value of $j$ where $N_{B}(n,x) < \nu$ (i.e., when
$j = n-|x|$), contribution (i) drops out.
At precisely the next value of $j$, $N_{B}(n,x) < \mu$
(i.e., $j = n-|x|+1$), both (i) and (ii) have no more
contributions.
Thus, both (i) and (ii) contribute up to $j=n-|x|-1$,
and then at $j=n-|x|$ there is a contribution from (ii)
alone. For larger values of $j$, there aren't any more contributions.

Recall that the problem of filling $n$ balls into $k$ urns,
such that no urn is empty, is the combinatoric problem of
strong composition,
and the number of ways are given
by ${n-1 \choose k-1}$. 

The total number of $n$-Feynman paths with $j$ switches is given by:

\begin{align}
\eta(n,x,j) &\equiv 
 {\frac{n+|x|}{2}-1 \choose
\lceil \frac{j+1}{2} \rceil-1 }
{\frac{n-|x|}{2}-1 \choose
\lfloor \frac{j+1}{2} \rfloor-1 }\nonumber\\
 &+
 {\frac{n-|x|}{2}-1 \choose
\lceil \frac{j+1}{2} \rceil-1 }
{\frac{n+|x|}{2}-1 \choose
\lfloor \frac{j+1}{2} \rfloor-1 },
\label{eq:gen}
\end{align}
assuming $j \le n -|x| -1$.
If $j = n -|x|$, then only the first summand in Eq. (\ref{eq:gen})
contributes.
The following identity can easily be verified:
\begin{equation}
{n \choose \frac{n\pm |x|}{2}} = \sum_{j=1}^{n-|x|-1} \eta(n,x,j)
 + 
 {\frac{n+|x|}{2}-1 \choose
\lceil \frac{n-|x|+1}{2} \rceil-1 }.
\label{eq:gen2}
\end{equation}
Recall that the LHS of Eq. (\ref{eq:gen2}) is same as the total number of Feynman paths obtained in Eq. (\ref{eq:TotFP}).

Note that in the case of equal forward
and backward paths, where we set $x=0$, the second summand
in the RHS of Eq. (\ref{eq:gen2}) vanishes.

Further, when a $B$ - string is concatenated with a $F$ (or vice - versa), an additional switch is introduced. For example, a string with two switches $BBFFB$, when concatenated with a $F$ to obtain the extended string $F*BBFFB$, has three switches. The additional switch so introduced in the extended needs to be accounted for. Let $\eta^*(n+1,x,j)$ denote the number of extended strings with $j$ switches, then:
\begin{align}
\eta^*(n+1,x,j)=\eta(n,x,j)+\eta(n,x,j-1),
\end{align} 
for $1<j<n-|x|$. For $j=1$ and $j=n-|x|+1$, we have:
\begin{equation}
\eta^*(n+1,x,j)=\eta(n,x,j).
\end{equation}

\begin{definition}[Parity]
Given an $n$ bit Feynman (extended) string, the parity $\epsilon$ takes the value $+1$ 
$($resp., $-1)$ when the number of $F \rightarrow F$ transitions in the string are even $($resp. odd$)$.
\end{definition} 

The parity of a Feynman extended string is obtained in the following lemma. 

\begin{lemma}
The parity $\epsilon(n,x,j)$ of Feynman $B$ or $F$- strings with $j$ switches is given by  
\begin{align}
\epsilon (n,x,j) = \left( -1 \right)^{\frac{n+x}{2} - \lfloor \frac{j+1}{2} \rfloor}
\label{eq: parityB}
\end{align}
and
\begin{align}
\epsilon (n,x,j)= (-1)^{\frac{n+x}{2} - \lceil \frac{j+1}{2} \rceil + 1} 
\label{eq: parityF}
\end{align}
respectively.
\label{lem: parity}
\end{lemma}

{\bf Proof.} Let $S$ be a $B$ - string with $j$ switches.  Cut the string at points where switches occur. This process creates $j+1$ sub-strings with $\lceil \frac{j+1}{2} \rceil$ $B$-type sub-strings and $\lfloor \frac{j+1}{2} \rfloor$, $F$-type sub-strings. Here, the $B$-type sub-strings do not alter the parity of the string $S$ and hence are ignored.  Distribute an $F$ bit to each of the $F$-type sub-strings as all sub-strings must be at least of size one. This leaves us with $ \frac{n+x}{2} - \lfloor \frac{j+1}{2} \rfloor$, $F$s which can then be redistributed among the sub-strings arbitrarily. Independent of the distribution of $F$s among the $F$-type sub-strings, the number of $F \rightarrow F$ transitions are unaltered when the sub-strings are conjoined. Hence, if $ \frac{n+x}{2} - \lfloor \frac{j+1}{2} \rfloor$ is even, $\epsilon=-1$, as there are odd number of $F \rightarrow F$ transitions, similarly, $\epsilon=1$ when $ \frac{n+x}{2} - \lfloor \frac{j+1}{2} \rfloor$ is odd.

The result can be proved analogously for $F$- strings.  

\hfill $\blacksquare$
\bigskip  

The above lemma also implies that the parity of a Feynman extended string is invariant under permutations which preserve the number of switches.

\begin{lemma}
\label{lem:coin}
A Feynman $B$ - string $($resp. $F$ - string$)$ with $j$ switches determines the coin state in an eigen basis $\ket{b,x}$: $b = j (mod \ 2)$ $(resp. b = j+1 (mod \ 2))$. 
\end{lemma}
{\bf Proof.} Let $S$ be a $B$ - string with $j$ switches, where $j$ is even. Cut the string at points where switches occur as described in lemma (\ref{lem: parity}). As $S$ is a $B$ - string, the first of the $j+1$ sub-strings contains only $B$s. This sub - string is followed by $F$ sub - string which in turn is followed by a $B$ sub-string and so on. It follows from this arrangement that the last sub-string contains $B$s. Every transition terminating with a $B$ results in the coin state $\ket{0}$ as evident from Fig. (\ref{fig:path}).  For an odd number of switches, $j$, in a $B$ - string, the above process of cutting and arranging the $j+1$ sub - strings, where $j+1$ is even, results in the $(j+1)^{th}$ sub-string to contain only $F$s. Every transition terminating with a $F$ results in the coin state $\ket{1}$ as evident from table Fig. (\ref{fig:path})).  

The result can be similarly shown for $F$ - strings.

\hfill $\blacksquare$
\bigskip

Hence,   using  the   above  results,   the  state   vector  component
corresponding to \textit{a given}  Feynman extended  string can expressly  be written
as:
\begin{align}
\ket{\xi(n,x,j,c)}= &  \epsilon(n+1,x,j)\cos^i\theta \sin^j\theta \ \times \nonumber \\
& \ket{\left(j+c\ (mod\ 2)\right),x},
\end{align} 
where, $c=0$ for $B$- strings and $c=1$ for $F$- strings.

We note that a $B$- string with $j$ switches and its dual have \textit{like} parity when $j$ is even and \textit{unlike} parity when $j$ is odd according to Eqs. (\ref{eq: parityB}) and (\ref{eq: parityF}). 

Defining the summing operations:
\begin{align}
\sum_1 &\equiv \sum_{x=-n}^n \dfrac{1+(-1)^{n+x}}{2} \sum_{m=1}^{\frac{n+|x|}{2}} 
\frac{\epsilon(n,x,2m)}{\eta^*(n+1,x,2m)} \nonumber\\
\sum_2 &\equiv \sum_{x=-n}^n \dfrac{1+(-1)^{n+x}}{2} \sum_{m=1}^{\frac{n+|x|+1}{2}}
\frac{\epsilon(n,x,2m-1)}{\eta^*(n+1,x,2m-1)},
\end{align}
we obtain the final state vector to be:
\begin{widetext}
\begin{align}
\ket{\psi(n)}&=\left(\sum_1 \alpha N_{B}(n,x)(\cos\theta)^{n-2m}\ (\sin\theta)^{2m} 
+ \sum_2 \beta e^{i\phi} N_{F}(n,x)(\cos\theta)^{n-2m+1}\ (\sin\theta)^{2m-1}\right) \ket{0,x} \nonumber\\
&+ \left(\sum_1 \beta e^{i\phi} N_{F}(n,x)(\cos\theta)^{n-2m}\ (\sin\theta)^{2m} 
- \sum_2 \alpha N_{B}(n,x)(\cos\theta)^{n-2m+1}\ (\sin\theta)^{2m-1}\right) \ket{1,x},
\end{align}
\end{widetext}

which may be considered as an explicit representation of a  QW parametrized by  position. This  allows us to  circumvent a direct calculation  of amplitudes at each  position. Its application will be explored elsewhere.

\section{Conclusion}
In this  work, a  closed form  expression was  obtained for  the state
vector corresponding  to arbitrary time step  for an $n$ -  step QW in
one dimension. The  Feynman path integral approach was  used to derive
all  possible   (backward  and   forward)  transition  rules   of  the
walker. The forward  and backward transitions were mapped  to a string
of  $F$s  and   $B$s  respectively.  A  recipe   for  calculating  the
probability  amplitude  for a  given  string  was provided  using  the
properties of these strings.

The formalism presented in  this work can be extended
  to  QW in  higher dimensions  including walks  by augmenting  the QW
  ``Feynman rules''  presented in  Fig. (\ref{fig:path}),  which would
  allow  us to  accommodate  QW in  higher dimensions  as  well QW  in
  arbitrary graphs. One can in principle extend this formalism to any
Markovian process evolving under a fixed unitary operator.

\bibliographystyle{unsrt}

\bibliography{comb} 

\end{document}